# Iris Image Processing in Compressive Sensing Scenario

(Student paper)

Radoje Darmanović
University of Montenegro

Tamara Bulatović
University of Montenegro

Seid Salković
University of Montenegro

*Abstract* — **This paper observes the application of the Compressive Sensing in reconstruction of the under-sampled iris images. Iris recognition represents form of biometric identification whose usage in real applications is growing. Compressive Sensing represents a novel form of sparse signal acquisition and recovering when small amount of data is a available. Different sparsity domains are considered and compared using various number of available image pixels. The theory is verified on iris images.**

*Keywords-iris recognition, compressed sensing*

## I. INTRODUCTION

In contrast to traditional forms of identity validation, biometrical are superior in a way that person doesn't need to have any additional items like personal ID card or passport, which is frequently lost or damaged. Biometrical data are also much more challenging to fake. Compared to other main forms of biometric recognition iris recognition has advantage of being contactless, eliminating stigma in some cultures [1]-[4]. Assumption is that iris is unique for every person, but because of nature of the problem, this statement can't be proven.

Iris recognition systems have found usage on mass scale in middle eastern countries and southern Asia in recent decade, and are increasingly applied to the other parts of the world. With greater number of citizens applying for biometric identity comes greater problem in form of rapidly increasing data accumulation which can represent a challenge for processing, transfer and storage. This is just one of many points where compressed sensing can be applied as solution as we will see.

Directive for standard bodies, International Organization for Standardization (ISO) and group that involves representatives from several different companies such as Registered Traveler Interoperability Consortium (RTIC) ember biometric data into smart cards, for purpose of avoiding using patented techniques into data formats and standards. If image is stored instead of templates that would result into almost thousand-fold increase in data size, which means there would be increased data transmission times and inability to ember the image data into allocated space in smart cards. Iris images in case of RTIC specification were only 400 bytes per eye. This means that compressibility and effects of lossy image compression on recognition performance have become critical. This begs the question: How much raw image data is really needed for iris recognition software to preform effectively? We'll discuss it in following texts.

## II. IRIS RECOGNITION

Iris recognition is the process of recognizing a person by analyzing the random pattern of the iris [1]-[4]. The iris structure and coloration are genetically linked, but the pattern details are not. Having in mind that an individual's irises are unique and structurally different, is enables an iris to be used in the process of person recognition.

Iris recognition is done on special terminals consisted of infrared camera. Person need to place one or both eyes on marked position. This conserves privacy as unwilling identifications proves impossible. Successful identification requires good quality image, without noise or blurred areas. Blur can easily occur in bad lighting conditions or in persons inability to stand still. Iris patterns are clearly seen in infrared while at the same time pigmentation, which can change over time, is removed.

Iris recognition technology combines computer vision, pattern recognition, statistical inference, and optics. First, the iris recognition algorithms were developed by John Daugman in 1990s with normalization and Gabor wavelets features as its main points. Iris recognition is done through several steps such as:
- Image capture
- Iris localization
- Normalization
- Feature extraction and template matching

### A. Image capture

Capturing device needs to ensure quality images for iris acquisition to be possible. Quality is reflected in image resolution, focus, noise and sharpness. To be able to pick up most details, subject needs to be close to capturing device, so concentric circles of pupil and iris can be clearly distinguishable [1]-[4].

### B. Iris localization

Iris localization is process of isolating iris region from the rest of image [1], [2]. This task can prove quite a challenge as eyelids and eyelashes can cover parts of iris, or iris and pupil diameters can have different radius ratios depending on the lighting condition. Localization is done by using integro-differential operator, given by equation [1], [2]:

$$max_{(r,x,y)} \left| G_0(r) * \frac{d}{dr} \oint_{r,x,y} \frac{I(x,y)}{2r\pi} dS \right| \quad (1)$$

*I(x,y)* represents the image containing an eye, *G* is Gaussian smoothing function, while *r* represents radius to search over an image. Result of the function is inner and outer iris boundary. Images with less then 50% of iris visible are rejected.

*C. Normalization*

Depending on the persons age, gender, etc. iris size can vary. To be able to search through iris codes efficiently, iris is transformed from polar to rectangular coordinates using Duagman sheet model [1], [2]. The centre of the pupil is considered as the reference point and a Remapping formula is used to convert the points on the Cartesian scale to the polar scale.

$$r' = \sqrt{\alpha}\beta \pm \sqrt{\alpha\beta^2 - \alpha - r1^2} \quad (2)$$
$$\alpha = \theta x^2 + \theta y^2 \quad (3)$$
$$\beta = \cos(\pi - \arctan\left(\frac{\theta y}{\theta x}\right) - \theta) \quad (4)$$

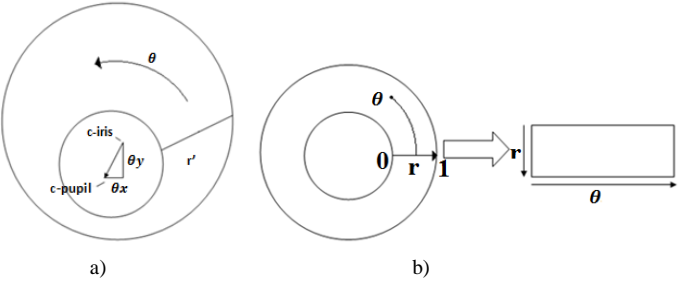

Figure 1: a) normalization process, b) unwrapping the iris

*D. Feature extraction*

To extract iris features, 2D Gabor filter are applied on textural features. Bidimensional Gabor is represented with the formula:

$$sgn_{\{Re,Im\}} \iint_{\mu,\varphi} I(\mu,\phi) e^{-i\omega(\theta-\varphi)} e^{-\frac{(r_0-\mu)^2}{r^2}} e^{-\frac{(\theta_0-\varphi^2)^2}{\beta^2}} \mu d\mu d\varphi \quad (5)$$

Where its result is complex valued bit whose result is either 1 or 0, depending on the sign of 2D integral. This way fixed bitmap is generated for the iris.
Templates are matched with the help of Hamming distance between two template bitmap patterns. Hamming distance for two binary templates Q and R can be calculated by:

$$HD = \frac{|(codeQ \otimes codeR) \cap maskQ \cap maskR|}{|maskQ \cap maskR|} \quad (6)$$

### III. COMPRESSED SENSING

Compressed sensing (CS) [5]-[25] uses a signal's inherent sparsity to allow for simultaneous data compression and acquisition. The standard process, based on Shannon-Nyquist sampling theorem, is to do data acquisition and compression sequentially. First, the data set is sampled at the Nyquist rate, and then it transformed into a sparse domain, that can be e.g. Discrete Cosine Transform (DCT), Discrete Fourier Transform (DFT), time-frequency domain, wavelet domain, etc. Then, compression is performed by removing all but the most important coefficients, i.e. only the coefficients with the largest amplitudes are kept. This can feel wasteful, since a large amount of data is firstly collected, and majority is discarded in the compression step. We may ask, can we collect less data but still reconstruct the full signal? In other words, can we perform compression during the sensing process instead of doing it afterwards [5], [8], [21].
Compressed sensing and sparsity offer the theory that allows us to do precisely that. If **X** is a signal in $\mathbb{R}^n$ and it is sparse in some domain **Ψ**, then the signal **X** can be reconstructed from *m* measurements where *m < n*. The sampling is done by using the matrix **Φ** that is incoherent with transform domain matrix **Ψ**. Reconstruction is done by using an optimization algorithms, which are based on recovering the signal **X** using smaller number of equations than unknowns.

$$X = \sum_{i=1}^{N} Si \, \Psi i = \Psi S \quad (7)$$

$$Y = \Phi X = \Phi\Psi S = \theta S \quad (8)$$

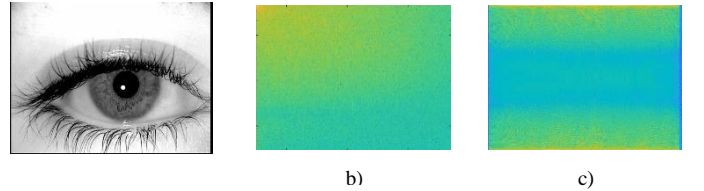

Figure 2: a) original image, b) DCT domain, and c) DFT domain

Since we are dealing here with the 2D data, it is important to note that there are no sparsity domain where images are strictrly sparse. Most images are considered as sparse in the DCT domain [8], [19]-[22], [25].
Numerical methods and algorithms that are used most for signal reconstruction are either $l_1$ minimization, greedy algorithms or total variation minimization. The Total-variation of image is based on variational parameters and is used in various image processing applications. Denoising and restoring noisy images would be an example of total-variation method. Let's say $x_n = x_0 + e$ is "noisy" observation of $x_0$, in that case we can restore $x_0$ by solving this minimization problem [8]:

$$\min_x TV(x) \; s.t. \; ||x_n - x||_2^2 < \varepsilon^2,$$

where $\varepsilon = ||\varepsilon||_2^2$ should hold and TV denotes the total-variation and could be approximated as [8], [21], [25]:

$$TV(x) = \sum_{i,j} ||D_{i,j}x||_2, D_{i,jx} = \begin{bmatrix} x(i+1,j) - x(i,j) \\ x(i,j+1) - x(i,j) \end{bmatrix}.$$

Total-variation methods that are based on denoising methods tend to eliminate noise without sacrificing detail or edges. This could be applied in compressive sensing to find an efficient reconstruction method, and for that goal we can say:

$$\min_x TV(x) \text{ s.t. } ||Ax - y||_2^2 < \varepsilon^2.$$

Total-variation reconstruction provides significantly better results when compared to $l_1$-minimization based reconstruction.

In iris recognition compressed sensing have several advantages:
- Through compressed sensing, we can take fewer measurements and reduce the power consumed by the sensor and speed up the data transfer and data processing.
- Raw output is encrypted and cannot be reconstructed without the measurement model which is very good for security
- Since each pixel is measured multiple times, compressed sensing is more robust than conventional imaging. For instance, in compressed sensing, if a fraction of the measurements are missing, the image can still be reconstructed.

## IV. EXPERIMENTAL RESULTS

Following results are obtained using TV compressed sensing algorithm in DCT and DFT domains using various number of sample measurements ranging from 10% to 40%.

Although very similar, DCT provides superior performance when image is sampled with lower number of samples. Table I represents Peak Signal-to-Noise Ratio values in dB, obtained from recovered images using different percent of the available pixels.

Table I: Peak Signal-to-Noise Ratio values in dB

|     | 10%   | 20%   | 30%   | 40%   |
|-----|-------|-------|-------|-------|
| DCT | 20.01 | 22.17 | 24.86 | 26.47 |
| DFT | 18.93 | 22.06 | 24.26 | 26.33 |

Reconstruction with fewer samples have noticeable loss of details which affected the iris recognition. In our case, recognition is proved to be impossible if less then 20% of the total number of samples is available in the DCT domain. For the DFT domain, the recognition is impossible if less then 30% is available, which is presented in Table II:

Table II: recognition is impossible if less then 20% (DCT) or 30% (DFT) of the total number of samples is available

|     | 10%  | 20%  | 30%  | 40%  |
|-----|------|------|------|------|
| DCT | FAIL | FAIL | PASS | PASS |
| DFT | FAIL | FAIL | FAIL | PASS |

For iris to be recognizes as valid, it needs to have hamming distance of less than 0.36. Hamming distances of images used in example are presented in next graph. Y axis represents hamming distance value, while X axis represents percent of samples.

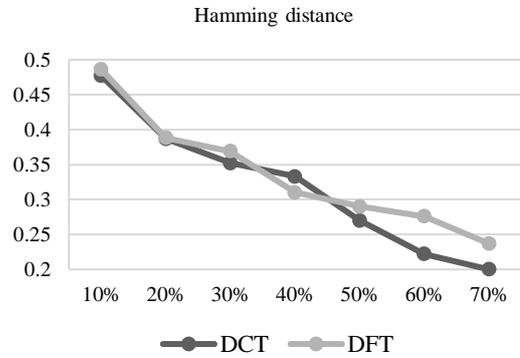

Figure 3: Hamming distances of images

To demonstrate robustness of iris recognition algorithm next two images represent visual difference between original image iris pixel map and image recovered using 30% of the samples for the DCT domain (which is necessary minimum for recognition):

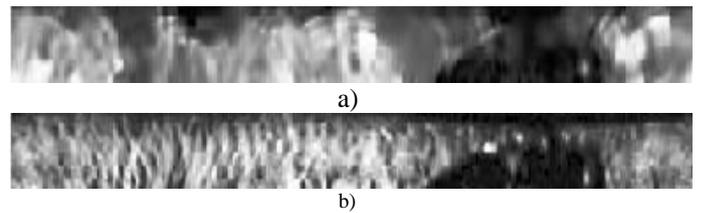

Figure 4: a) original image iris pixel map b) image recovered using 30% of the samples for DCT domain

Low quality or blurry images suffer from yet another problem, pupil and iris boundaries are hard, sometimes impossible to find which means that recognition is not possible. The images recovered using different sparsity domains and different number of available samples are shown in Fig. 5.

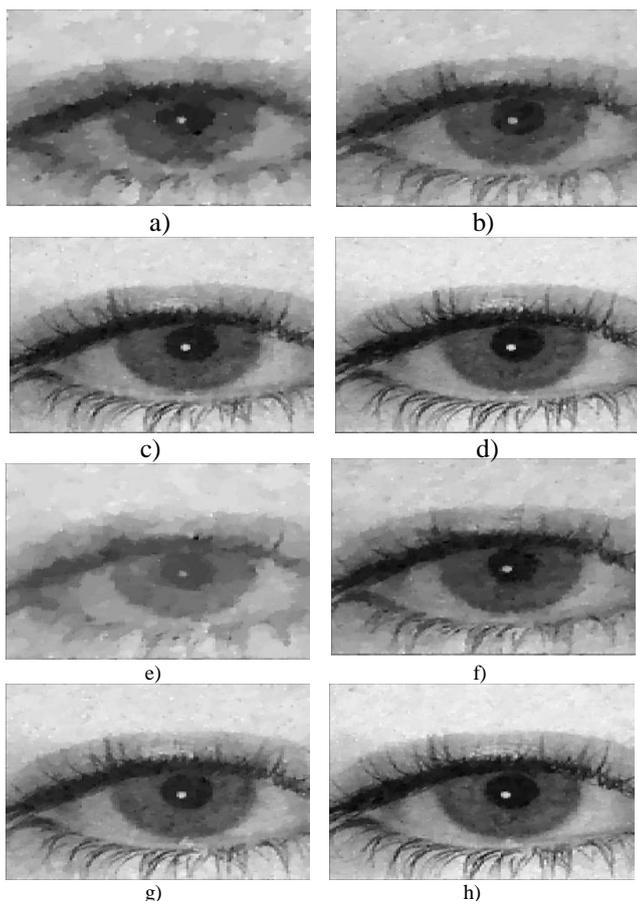

Figure 5: a) DCT 10%,, b) DCT 20%, c) DCT 30%, d) DCT 40%, e) DFT 10%, f) DFT 20%, g) DFT 30%, h) DFT 40%

## V. CONCLUSION

The CS application in iris recognition is tested in the paper. Different sparsity domains are observed – the DFT and the DCT domain. Also, CS image reconstruction is tested using different number of samples available. The DCT domain provides better reconstruction and consequently successful reconstruction using smaller number of available pixels compared to the DFT (20% for the DCT, while number of available pixels should be at least 30% for the DFT). This paper can be extended in a way of using additional domains like wavelets.